\documentclass[aps,prd,groupedaddress,showpacs,showkeys,nofootinbib]{revtex4}%
\usepackage{epsfig,amssymb,amsmath}
\usepackage[english]{babel}%
\usepackage{amsmath}%
\usepackage{amsfonts}%
\usepackage{amssymb}%
\usepackage{graphicx}

\usepackage{graphicx}
\usepackage{dcolumn}
\usepackage{bm}
\usepackage{epsfig}
\begin{document}
\preprint{test1}
\title{ Scalar-tensor cosmology with  $R^{-1}$ curvature correction by Noether Symmetry}
\author{ H. Motavali${}^{\flat}$, S.Capozziello${}^{\sharp}$\footnote{Corresponding author: capozziello@na.infn.it}, M. Rowshan Almeh Jog${}%
^{\flat}$}   \affiliation{${}^{\flat}$ Faculty of Physics,
University of Tabriz, Tabriz 51664,
Iran,}\affiliation{${}^{\sharp}$ Dipartimento di Scienze Fisiche,
Universit\`{a} di Napoli "Federico II" and INFN Sez. di Napoli,
Compl. Univ. Monte S. Angelo, Ed.N, Via Cinthia, I-80126 Napoli,
Italy.}

\date{\today}

\begin{abstract}
We discuss  scalar-tensor cosmology with an extra $R^{-1}$
correction by the Noether Symmetry Approach. The existence of such
a symmetry selects the forms of the coupling $\omega(\varphi)$, of
the potential $V(\varphi)$ and allows to obtain physically
interesting exact cosmological solutions.
 \end{abstract}

\maketitle
\section{Introduction}
Recent observational data  indicate that $\simeq 70\%$ of the
today cosmological energy density  is dominated by some form of
"dark energy" which can be described, in the simplest way, by the
 cosmological constant $\Lambda$
\cite{Perl97,perlal,rei&al98,Riess00}. Such an ingredient should
explain the accelerated expansion of the observed universe,
firstly deduced by luminosity distance measurements of SNeIa
supernovae. However, even though the presence of a dark energy
component is appealing in order to fit observational results with
theoretical predictions, its fundamental nature still remains an
open question.

Although several models describing the dark energy component have
been proposed over the past few years, one of the first physical
realizations of quintessence was a cosmological scalar field,
which dynamically induces a repulsive gravitational force, causing
an accelerated expansion of the universe.

The existence of such a large proportion of dark energy in the
universe presents a large number of theoretical problems. Firstly,
why do we observe the universe at exactly the time in its history
when the vacuum energy dominates over matter (this is known as the
\textit{cosmic coincidence} problem). The second issue, which can
be thought of as a \textit{fine tuning problem}, arises from the
fact that if the vacuum energy is constant, like in the pure
cosmological constant scenario, then at the beginning of the
radiation era the energy density of the scalar field should have
been vanishingly small with respect to the radiation and matter
components. This poses the problem that, in order to explain the
inflationary behavior of the early universe and the late time dark
energy dominated regime, the vacuum energy should evolve and
should not  simply be a \textit{constant}.

Some recent works have shown that the fine-tuning problem could be
alleviated by selecting a subclass of quintessence models, which
admit a \textit{tracking behavior} \cite{stein2}, and in fact, to
a large extent, the study of scalar field quintessence cosmology
is often limited to such a subset of solutions. In scalar field
quintessence, the existence condition for a tracker solution
provides a sort of selection rule for the potential $V(\phi)$ (see
\cite{all03} for a critical treatment of this question), which
should somehow arise from a high energy physics mechanism (the so
called \textit{model building problem}). Also, adopting a
phenomenological point of view, where the functional form of the
potential $V(\phi)$ can be determined from observational
cosmological functions, for example the luminosity distance, we
still cannot avoid a number of problems. For example, an attempt
to reconstruct the potential from observational data (and also
fitting the existing data with a linear equation of state) shows
that a violation of the weak energy condition (WEC) is not
completely excluded \cite{cald}, and this would imply a
\textit{superquintessence regime}, during which $w_{\phi}<-1$
(\textit{phantom regime}). However, it turns out that assuming a
dark energy component with an arbitrary scalar field Lagrangian,
the transition from regimes with $w_{\phi} \geq -1$ to those with
$w_{\phi}<-1$ (i.e. crossing the so called \textit{phantom
divide}) are probably physically impossible since they are either
described by a discrete set of trajectories in the phase space or
are unstable \cite{hu,vik}. These shortcomings have been recently
overcome by considering the \textit{unified phantom cosmology}
\cite{odintsov} which, by taking a generalized scalar field
kinetic sector into account, allows one to achieve models with
natural transitions between inflation, dark matter, and dark
energy regimes. Moreover, in recent works, a dark energy component
has been modelled also in the framework of scalar-tensor theories
of gravity, also called extended quintessence (see for instance
\cite{star,fuji2,francesca1,DGP1,chiba,uzan,copeland,metric,palatini,GRGrev}).

It turns out that they are compatible with a \textit{peculiar}
equation of state $w\leq -1$, and provide a possible link to
issues occurring in non-Newtonian gravity \cite{fuji2}. In such
theoretical backgrounds, the accelerated expansion of the universe
is due to the effect of the non-standard form of the gravitational
action. In extended quintessence cosmologies (EQ) the scalar field
is coupled to the Ricci scalar $R$ in the Lagrangian density  of
the theory: the standard term  $ \displaystyle 16 \pi
G_{\ast}\,\,R$ is replaced by $\displaystyle 16 \pi F(\phi)\,\,
R$, where $F(\phi)$ is a function of the scalar field, and
$G_{\ast}$ is the {\it bare} gravitational constant, generally
different from the Newtonian constant $G_N$ measured in
Cavendish-type experiments \cite{star}.

Of course, the coupling is not arbitrary, but it is subjected to
several constraints, mainly arising from the time variation of the
constants of nature \cite{uzan2}. In EQ models, a scalar field has
indeed a double role: it determines at any time the effective
gravitational constant and contributes to the dark energy density,
allowing some different features from the minimally coupled case
\cite{uzan2}. Actually, while in the framework of the minimally
coupled theory we have to deal with a fully relativistic
component, which becomes homogeneous on scales smaller than the
horizon, so that standard quintessence cannot cluster on such
scales, in the context of non-minimally coupled quintessence
theories the situation is different, and the scalar field density
perturbations behave like the perturbations of the dominant
component at any time. This is referred to as  {\it gravitational
dragging} (\cite{francesca1}).

On the other hand, the cosmic speed up can be simply explained
considering some sub-dominant terms of geometric origin like
$R^{-1}$, where $R$ is the Ricci curvature scalar,  which becomes
dominant toward small curvature regimes (see e.g. \cite{metric}).
In fact, it is possible to show that, by adding these terms  to
the Hilbert-Einstein action and varying  with respect to the
metric, such modified field equations  naturally produce the
observed cosmological acceleration. The simplest action of these
models is:
\begin{eqnarray}
S=\frac{1}{8\pi
G_N}{\int\left(R-\frac{\mu_{0}^{4}}{R}\right)\sqrt{-g}\;d^{4}x}
\end{eqnarray}
where  $G_N$ is the Newton's gravitational constant and $\mu_{0}$
is a constant. The Palatini variation of this action is studied,
for example, in Ref. \cite{palatini}. However, we need some
additional ingredient to fit  the observed data and physical
constraints at every redshift so a modified scalar-tensor theory
could be a more suitable candidate  to achieve the whole observed
dynamics. In this perspective, the investigation of theories like
\begin{eqnarray}
\label{nonminimal}
S=\int\left[\phi\left(R-\frac{\mu_{0}^{4}}{R}\right)+\frac{\omega(\phi)}{\phi}g^{\mu\nu}\nabla
_{\mu}\phi\nabla_{\nu}\phi-V(\phi)\right]\sqrt{-g} d^{4}x
\end{eqnarray}
is in order.  Here $\phi$ denotes a real scalar field,
non-minimally coupled to gravity while $\omega(\phi)$ and
$V(\phi)$ are the coupling function and a self-interacting
potential, respectively.

A physical criterion to achieve general cosmological solutions
could be by the \textit{Noether Symmetry Approach}  which revealed
a useful tool to fix the forms of the coupling and the potential
\cite{22}, and, very recently, also the form of $f(R)$
\cite{prado,antonio}. From a mathematical point of view, the
method lies on the fact the presence of symmetries  selects cyclic
variables which allow to reduce the dynamics and then to integrate
the equations of motion. From a physical point of view, any
Noether symmetry is associated to some conserved quantity. This
fact allows to select physically viable models (see for example
\cite{22}) and constitutes a criterion to select suitable
effective Lagrangian (in particular, the forms of the coupling, of
the self-interacting potential and the higher-order corrections).

Specifically, in this letter, we work out the above action
(\ref{nonminimal}) searching for  Noether Symmetries in order to
see if the coupling, the self-interacting potential and the
$R^{-1}$ can be related in physically viable models. Besides, as
we will see, this procedure allows to exactly integrate the
equations of motion.

The letter is organized as follows. In Sect. II, we search for
Noether symmetries for the above action selecting the coupling and
the potential. Sect. III is devoted to find out the cosmological
solutions and the discussion of the various sub-cases. Concluding
remarks and conclusions are reported in Sect. IV.

\section{The Noether symmetry }
In order to search for Noether Symmetries, it is convenient to
recast the action (\ref{nonminimal}) by redefining
$\phi=\varphi^{2}$ and $\mu_{0}^4=-\mu$, that is
\begin{eqnarray}
S=\int\left[\varphi^{2}\left(R+\frac{\mu}{R}\right)+4\omega(\varphi)g^{\mu\nu}\nabla
_{\mu}\varphi\nabla_{\nu}\varphi-V(\varphi)\right]\sqrt{-g}d^{4}x
\end{eqnarray}
\\
Using the Friedmann-Robertson-Walker (FRW) metric, the scalar
curvature takes the form ${\displaystyle
R=6\left(\frac{\ddot{a}}{a}+\frac{\dot{a}^{2}}{a^{2}}+\frac{k}{a^{2}}\right)}$,
where $a(t)$ is the scale factor of the universe and the dot
denotes the derivative with respect to time, with $k=\pm1,0$ .
Now, using the Lagrange multipliers method \cite{lambiase}, one
can rewrite the action (3) as follows
\\
\begin{eqnarray}
S=\int\left[\varphi^{2}\left(R+\frac{\mu}{R}\right)+4\dot{\varphi}^{2}\omega(\varphi)-V(\varphi)+\lambda_{1}
(R-6\left(\frac{\ddot{a}}{a}+\frac{\dot{a}^{2}}{a^{2}}+\frac{k}{a^{2}}\right)\right]\sqrt{-g}d^{4}x
\end{eqnarray}
\\
where, the scalar curvature $R$ and scale factor $a$ are
considered as two independent variables, and $\lambda_{1}$ is a
Lagrange multiplier. The parameter $\lambda_{1}$ can be determined
by varying the action with respect to $R$ , that is
\\
\begin{eqnarray}
\lambda_{1}=\varphi^{2}(\mu R^{-2}-1)
\end{eqnarray}
\\
Now, in order to apply the Noether symmetry approach, one can
easily show that,in a $FRW$ manifold, the Lagrangian  related to
the action (4) takes the point-like form \cite{22}
\\
\begin{eqnarray}
{\cal L}=2a^{3}\varphi^{2}\mu q+6(\mu
q^{2}-1)(2a^{2}\varphi\dot{a}\dot{\varphi}+\varphi^{2}
\dot{a}^{2}a)+12\mu\varphi^{2}a^{2}q\dot{a}\dot{q}-6\varphi^{2}k
a(\mu q^{2}-1)+a^{3}(4\omega(\varphi)\dot{\varphi}^{2}-V(\varphi))
\end{eqnarray}
\\
where $q=R^{-1}$ is a new variable. This means that we are
considering an effective theory with \textit{two} scalar fields.
The corresponding Euler-Lagrange equations are given by
\\
\begin{eqnarray}
(\mu q^{2}-1)\left(2\varphi H^{2}+\frac{\varphi
k}{a^{2}}+\dot{H}\varphi\right)+\frac{1}{3}\frac{d\omega}{d\varphi}\dot{\varphi}^{2}+
\frac{2}{3}\omega(\varphi)\ddot{\varphi}-\frac{1}{3}\mu
q\varphi+2\omega(\varphi)\dot{\varphi}H+\frac{1}{12}\frac{dV}{d\varphi}=0
\end{eqnarray}
\\
\begin{eqnarray}
 (\mu q^{2}-1)\left[2\varphi\dot{\varphi}H+\varphi^{2}(\frac{3}{2}H^{2}+\frac{k}{2a^{2}}+\dot{H}
+\frac{d(\varphi\dot{\varphi})}{dt})\right]+\frac{1}{4}V(\varphi)-\omega(\varphi)\dot{\varphi}^{2}
+\mu\varphi^{2}\frac{d(q\dot{q})}{dt}+2\mu\varphi
q\dot{q}(2\dot{\varphi}+H\varphi)=0
\end{eqnarray}
with the constraint, derived from the definition of the scalar
curvature $R$,
\begin{eqnarray}
6\left(2H^{2}+\dot{H}+\frac{k}{a^{2}}\right)=\frac{1}{q}
\end{eqnarray}
Here ${\displaystyle H=\frac{\dot{a}}{a}}$ is  the Hubble
parameter. Eqs. (7) and (8) are equivalent to the second order
Einstein equation and to the Klein-Gordon equation, respectively.
Finally, one can choose the initial conditions of these field
equations such that the energy function associated with the
Lagrangian (6) vanishes, that is
\\
\begin{eqnarray}
E_{\cal L}=\dot{a}\frac{\partial {\cal L}}{\partial
\dot{a}}+\dot{q}\frac{\partial {\cal L}}{\partial
\dot{q}}+\dot{\varphi}\frac{\partial {\cal L}}{\partial
\dot{\varphi}}-{\cal L}=0\,,
\end{eqnarray}
\\
or explicitly
\begin{eqnarray}
(\mu
q^{2}-1)\left(\varphi\dot{\varphi}H+\frac{1}{2}\varphi^{2}H^{2}+\frac{\varphi^{2}k}{2a^{2}}\right)+\frac{1}{3}
\omega(\varphi)\dot{\varphi}^{2}+\frac{1}{12}V(\varphi)+\mu
\varphi^{2}H q\dot{q}-\frac{1}{6}\mu q \varphi^{2}=0
\end{eqnarray}
\\
which corresponds to the $\{0,0\}$ Einstein equation. Now, let us
introduce the \textit{lift vector field} $X$  \cite{marmo} as an
infinitesimal generator of the Noether symmetry in the tangent
space $TQ\{a,\dot{a},\varphi,\dot{\varphi},q,\dot{q}\}$ related to
the configuration space $Q=\{a,q,\varphi\}$ as follows
\\
\begin{eqnarray}
X=A\frac{\partial}{\partial a}+B\frac{\partial}{\partial
\varphi}+C\frac{\partial}{\partial
q}+\dot{A}\frac{\partial}{\partial\dot{a}}+\dot{B}\frac{\partial}{\partial\dot{\varphi}}+
\dot{C}\frac{\partial}{\partial\dot{q}}
\end{eqnarray}
\\
where $A,B$ and $C$ are unknown functions of the variables
$a,\varphi$ and $q$. The existence of Noether symmetry for the
dynamics implies that the vector field $X$ is non-trivial and the
Lie derivative of the Lagrangian, with respect to this vector
field, vanishes
\begin{eqnarray}
\nonumber
 L_{X}{\cal L}=0
\end{eqnarray}
\\
Explicitly, this condition leads to the following differential
equations
\begin{eqnarray}
\nonumber 6a^{2}\mu \varphi^{2}q
A-3a^{2}V(\varphi)A-Ba^{3}\frac{dV}{d\varphi}+4\mu a^{3}\varphi q
B+
\end{eqnarray}
\begin{eqnarray}
+2a^{3}\varphi^{2}\mu C+6 k \varphi^{2}A(1-\mu q^{2})+12 k a
\varphi B(1-\mu q^{2})-12 \mu k a \varphi^{2}q C=0
\end{eqnarray}
\begin{eqnarray}
3\omega(\varphi)A+Ba\frac{d\omega}{d\varphi}+3(\mu
q^{2}-1)\varphi\frac{\partial A}{\partial
\varphi}+2\omega(\varphi)a\frac{\partial B}{\partial\varphi}=0
\end{eqnarray}
\begin{eqnarray}
(\mu q^{2}-1)\left(\varphi A+2Ba+2a \varphi\frac{\partial
A}{\partial a}+2a^{2}\frac{\partial B}{\partial a}\right)+2\mu q
a\varphi\left(C+a\frac{\partial C}{\partial a}\right)=0
\end{eqnarray}
\begin{eqnarray}
(\mu q^{2}-1)\left(2\varphi A+Ba+a\varphi\frac{\partial
A}{\partial a}+\varphi^{2}\frac{\partial
A}{\partial\varphi}+\varphi a\frac{\partial B}{\partial
\varphi}\right)+\frac{2}{3}\omega(\varphi)a^{2}\frac{\partial
B}{\partial a}+\mu qa\varphi\left(2C+\varphi\frac{\partial
C}{\partial \varphi}\right)=0
\end{eqnarray}
\begin{eqnarray}
(\mu q^{2}-1)\left(\varphi\frac{\partial A}{\partial
q}+a\frac{\partial B}{\partial q}\right)+2\mu q A\varphi+2\mu
qaB+\mu a\varphi\left(C+q\frac{\partial A}{\partial
a}+q\frac{\partial C}{\partial q}\right)=0
\end{eqnarray}
\begin{eqnarray}
\mu q\varphi^{2}\frac{\partial
A}{\partial\varphi}+\frac{2}{3}\omega(\varphi)a\frac{\partial
B}{\partial q}+\varphi(\mu q^{2}-1)\frac{\partial A}{\partial q}=0
\end{eqnarray}
and
\begin{eqnarray}
\frac{\partial A}{\partial q}=0
\end{eqnarray}
Putting (19) into (18) implies
\begin{eqnarray}
3\mu q\varphi^{2}\frac{\partial
A}{\partial\varphi}+2\omega(\varphi)a\frac{\partial B}{\partial
q}=0
\end{eqnarray}
By choosing $A=A_{0}a^{n}\varphi^{m},B=B_{0}(q)a^{l}\varphi^{s}$
and substituting them into Eq. (20), we get
\begin{eqnarray}
B_{0}(q)=-\frac{3}{4}\mu\frac{mA_{0}}{\omega_{0}}q^{2}+k_{1}
\end{eqnarray}
\begin{eqnarray}
\omega(\varphi)=\omega_{0}\varphi^{m-s+1}
\end{eqnarray}
where $A_{0},\omega_{0}$ and $k_{1}$ are constant. By substituting
this results into (14) we get
\begin{eqnarray}
\omega_{0}=m=1\;\;\;\;\mbox{and} \;\;\;\;s=2
\end{eqnarray}
Taking into account Eqs. (21), (22) and (23), we get the solutions
\begin{eqnarray}
\omega(\varphi)=1
\end{eqnarray}
\begin{eqnarray}
A=A_{0}a^{n}\varphi
\end{eqnarray}
\begin{eqnarray}
 B=(-\frac{3}{4}\mu A_{0}q^{2}+k_{1})a^{n-1}\varphi^{2}
\end{eqnarray}
An important remark is in order at this point. In the case
$\mu=0$, such solutions are ruled out, if $\varphi$ is massless,
by gravity tests on Solar System. This is not true for $\mu\neq
0$. In this case, the previously mentioned tests strongly
constrain the allowed masses for $\varphi$ and therefore the
parameters in the potential $V(\varphi)$. For a detailed
discussion on the effective scalar field mass constrained by Solar
System tests see \cite{odi,tsu,tsuio}.

In view of these solutions, Eqs. (15), (16) and (17) read
\begin{eqnarray}
\nonumber
\left[\left(\frac{7}{2}+\frac{2k_{1}}{A_{0}}+n\right)-3\mu
q^{2}\right]a^{n}\varphi^{2}\mu A_{0}q+\mu
a\varphi\left(C+q\frac{\partial C}{\partial q}\right)=0
\end{eqnarray}
\begin{eqnarray}
\nonumber (\mu
q^{2}-1)A_{0}a^{n}\varphi^{2}\left[(3+n+\frac{k_{1}}{A_{0}})-\frac{9}{4}\mu
q^{2}\right]+\frac{(1-n)}{2}\mu
q^{2}A_{0}a^{n}\varphi^{2}+a\varphi \mu
q\left(2C+\varphi\frac{\partial C}{\partial\varphi}\right)=0
\end{eqnarray}
and
\begin{eqnarray}
\nonumber (\mu
q^{2}-1)A_{0}a^{n}\varphi^{2}\left[\left(1+2n+\frac{2k_{1}}{A_{0}}\right)-\frac{3}{2}n\mu
q^{2}\right]+2\mu qq\varphi\left(C+a\frac{\partial C}{\partial
a}\right)=0
\end{eqnarray}
These equations are satisfied if
\begin{eqnarray}
q^{2}=q_{0}^{2}=\frac{1}{\mu}G
\end{eqnarray}
\begin{eqnarray}
C=f_{0}a^{n-1}\varphi
\end{eqnarray}
\begin{eqnarray}
\nonumber
f_{0}=\frac{\beta_{0}}{q_{0}}+\frac{A_{0}q_{0}}{4}\left(3\mu
q_{0}^{2}-7-2n-\frac{4k_{1}}{A_{0}}\right)
\end{eqnarray}
\begin{eqnarray}
\nonumber
G=\frac{2n-\frac{3}{n}+\frac{2k_{1}}{A_{0}}(1-\frac{3}{n})}{n-\frac{3}{n}+1+\frac{2k_{1}}{A_{0}}(1-\frac{3}{n})}
\end{eqnarray}
\begin{eqnarray}
\nonumber
\beta_{0}=\frac{A_{0}}{2\mu}\frac{2n(n^{2}+n-1)-5+\frac{k_{1}}{A_{0}}(6n^{2}-4n-16)+
\frac{k_{1}^{2}}{A_{0}^{2}}(4n-12)}{n^{2}+n-3+\frac{2k_{1}}{A_{0}}(n-2)}
\end{eqnarray}
where $G,f_{0},q_{0}$ and $\beta_{0}$ are constant. In conclusion,
the Noether symmetry for the Lagrangian (6) exists and the vector
field $X$ is determined by (25), (26) and (27) and (28) while the
functional form of $\omega(\varphi)$ is given by (24).

It is straightforward to obtain a general self-interaction
potential from Eq. (13) as
\begin{eqnarray}
V(\varphi)=\lambda
\varphi^{2}+k_{2}\varphi^{\frac{1}{\Lambda_{2}}}
\end{eqnarray}
where, we have used the definitions
\begin{eqnarray}
\nonumber  k_{2}=\frac{1}{12}\left(9A_{0}\mu
q_{0}^{2}-6A_{0}+\frac{12\mu f_{0}}{1-\mu q_{0}^{2}}\right)
\end{eqnarray}
\begin{eqnarray}
\nonumber  \lambda=\frac{\Lambda_{1}}{1-2\Lambda_{2}}
\end{eqnarray}
\begin{eqnarray}
\Lambda_{1}=\left(2q_{0}-3q_{0}^{3}+\frac{2f_{0}}{3A_{0}}\right)\mu
\end{eqnarray}
\begin{eqnarray}
\nonumber \Lambda_{2}=\frac{1}{4}\mu
q_{0}^{2}-\frac{k_{1}}{3A_{0}}
\end{eqnarray}

The existence of the Noether symmetry means that there exists a
constant of motion. In this case, the conserved quantity
corresponding to the Noether symmetry can be obtained using the
Cartan one-form associated with the Lagrangian (6), that is
\begin{eqnarray}
\nonumber \theta_{\cal L}=\frac{\partial {\cal L}}{\partial
\dot{a}}da+\frac{\partial {\cal L}}{\partial
\dot{\varphi}}d\varphi+\frac{\partial {\cal L}}{\partial
\dot{q}}dq
\end{eqnarray}
By contracting $\theta_{\cal L}$ with $X$ one obtains the
following required constant of motion
\begin{eqnarray}
\nonumber F_{0}=i_{X}\theta_{\cal L}=A_{0}\varphi a^{n}((\mu
q^{2}-1)(12a^{2}\varphi\dot{\varphi}+12\varphi^{2}a\dot{a}+72\mu
a^{2}\varphi^{2}q\dot{q})+
\end{eqnarray}
\begin{eqnarray}
+12f_{0}a^{n+1}\varphi^{3}\mu\dot{a}q+a^{n-1}\varphi^{2}(k_{1}-\frac{3}{4}\mu
A_{0}q_{0}^{2})(12(\mu
q^{2}-1)a^{2}\dot{a}\varphi+8a^{3}\omega(\varphi)\dot{\varphi})
\end{eqnarray}

\section{ The cosmological solution}
Starting from (9) and (27), it is straightforward to get the
following general solution for the scale factor
\begin{eqnarray}
\nonumber
a(t)=\sqrt{6kq_{0}+\alpha_{1}\exp(\frac{t}{\sqrt{3q_{0}}})+\alpha_{2}\exp(\frac{-t}{\sqrt{3q_{0}}})}
\end{eqnarray}
 where $\alpha_{1}$ and $\alpha_{2}$ are arbitrary constants. In special case,
 by choosing $k=\alpha_{2}=0$ and $\alpha_{1}=a_{0}^{2}$, this solution takes the standard de Sitter form
\begin{eqnarray}
a(t)=a_{0}\exp(\alpha t)
\end{eqnarray}
where ${\displaystyle \alpha=\frac{1}{2\sqrt{3q_{0}}}}$. Clearly
this is a singularity free solution evolving as an hyperbolic
cosine. It shows  accelerated phases for $t\rightarrow\pm\infty$
so  both inflationary and dark energy behaviors are easily
achieved.

Some interesting sub-cases can be obtained considering the field
potential (29). For $k_{2}=0$, it takes the form
$V(\varphi)=\lambda\varphi^{2}$. In addition, one can use the
constant of motion (31) and the scale factor (32) to find a
solution for $\varphi(t)$. To this purpose, we rewrite (31) as
\begin{eqnarray}
F_{0}=a^{n+2}(A_{0}(\mu
q_{0}^{2}-1)(12\varphi^{2}\dot{\varphi}+12\varphi^{3}\alpha)+12f_{0}\varphi^{3}\mu\alpha
q_{0}+\varphi^{2}(k_{1}-\frac{3}{4}\mu A_{0}q_{0}^{2})(12(\mu
q_{0}^{2}-1)\alpha\varphi+8\dot{\varphi}))
\end{eqnarray}
and then
\begin{eqnarray}
\varphi(t)=\varphi_{0}\exp(-\vartheta_{0} t)
\end{eqnarray}
with
\begin{eqnarray}
\nonumber
\vartheta_{0}=\frac{(n+2)\alpha}{3},\;\;\;\varphi_{0}=\left(\frac{F_{0}}{u_{0}}\right)^{\frac{1}{3}}
\end{eqnarray}
and
\begin{eqnarray}
\nonumber u_{0}=a_{0}^{n+2}(12A_{0}(\mu
q_{0}^{2}-1)(12\vartheta_{0}+12\alpha)+12f_{0}\mu\alpha
q_{0}+(k_{1}-\frac{3}{4}\mu
 q_{0}^{2}A_{0})(12\alpha(\mu q_{0}^{2}-1)+8\vartheta_{0}))
\end{eqnarray}
It must be stressed that these results have been obtained due to
the existence of the Noether symmetry, and one can easily check
that these solutions are consistent with the corresponding field
equations. In this case ($k_{2}=0$), solutions (32) and (34)
satisfy the Eqs. (7) and (8) which now assume the forms
\begin{eqnarray}
\nonumber 2\varphi H^{2}(\mu
q_{0}^{2}-1)+\frac{2}{3}\ddot{\varphi}-\frac{1}{3}\mu
 q_{0}\varphi+2\dot{\varphi}H+\frac{\lambda}{6}\varphi=0
\end{eqnarray}
\begin{eqnarray}
 \nonumber (\mu
q_{0}^{2}-1)\left(2\varphi\dot{\varphi}H+\frac{3}{2}H^{2}\varphi^{2}+\frac{d(\varphi\dot{\varphi})}{dt}\right)+\frac{1}{4}\lambda\varphi^{2}-\dot{\varphi}^{2}=0
\end{eqnarray}
respectively, with the following definitions of the constants
\begin{eqnarray}
q_{0}^{2}=\frac{1}{\mu}\frac{348n+108n^{2}+16n^{3}+473}{(4n+4n^{2}+73)(13+2n)}
\end{eqnarray}
\begin{eqnarray}
\nonumber
\lambda=\frac{583n+327n^{2}+16n^{3}-4n^{4}+1508}{27q_{0}(4n+4n^{2}+73)}
\end{eqnarray}
Clearly $n$ is a free parameter depending on the constant of
motion.

Another interesting case is for $\lambda=0$ and
$\Lambda_{2}=\frac{1}{4}$.  The self-interaction potential takes
the form $V(\varphi)=k_{2}\varphi^{4}$. As it is well known, this
potential is widely used in the discussion of vibrations of
polyatomic molecules \cite{17} and it is widely used in chaotic
inflationary models also if, strictly speaking,  the
$V(\varphi)\sim \varphi^{2}$ potential is the prototype of chaotic
inflationary potentials \cite{22,23,24}. However, we have to say
that quartic $\varphi^{4}$ potentials are almost ruled out by
current observations (see, for instance, \cite{kinney}).

In this case, the general solution of the equations of motion is
achieved  numerically while particular solutions have a power-law
form as discussed in \cite{pla} for non-minimally coupled theories
without the $R^{-1}$ correction.

\section{Concluding remarks}
We have explored the conditions for the existence of a Noether
symmetry in a scalar-tensor theory of gravity, with an extra
$R^{-1}$ term, in which the coupling function and the generic
potential are unknown. The  motivation for this study is  that we
want to construct cosmological models capable of achieving
inflationary and dark energy phases. To this goal, we need two
fields leading the two eras which, in our case, are $\varphi$ and
$q=R^{-1}$.

We have shown that the existence of the symmetry fixes the
coupling and the self-interacting potential which have physically
interesting forms. Furthermore, it allows to achieve exact
cosmological solutions which are singularity free and suitable to
mimic inflationary and dark energy behaviors.

A more physically appealing model should consider the role of
standard perfect fluid matter and should fit also the dust
dominated phase \cite{prado}, but in this cases the Noether
symmetry cannot always achieved.

However, also if we have considered a phenomenological model, the
important lesson of this research is that, as shown also in other
contexts  \cite{22,antonio}, the Noether symmetry is a powerful
approach to select physically motivated solutions.

\end{document}